\documentclass[preprint,12pt,sort,authoryear]{elsarticle}

\usepackage{amssymb}
\usepackage{amsmath}
\usepackage{tabularx} 
\usepackage{geometry} %
\usepackage{booktabs}   
\usepackage{multirow}   
\usepackage{hyperref}
\usepackage{makecell}
\usepackage{xcolor} 
\usepackage{colortbl} 
\usepackage{natbib}
\definecolor{customblue}{RGB}{31,119,180}
\definecolor{customgreen}{RGB}{152,223,138}
\definecolor{custombrown}{RGB}{140,86,75}
\definecolor{customgray}{RGB}{199,199,199}
\journal{Computers \& Education}

\begin{document}

\title{Mapping the AI Divide in Undergraduate Education: Community Detection in Disciplinary Networks and Survey Evidence}

\author[a,b]{Liwen Zhang}
\ead{502023110038@smail.nju.edu.cn}
\ead[url]{https://orcid.org/0009-0005-6003-4717}

\author[a,b]{Wei Si}
\ead{221810021@smail.nju.edu.cn}

\author[a,b]{Ke-ke Shang\corref{cor1}} 
\ead{kekeshang@nju.edu.cn}
\ead[url]{https://orcid.org/0000-0002-7454-4276}

\author[b]{Jiangli Zhu}
\ead{zhujl@nju.edu.cn}

\author[b]{Xiaomin Ji\corref{cor1}}  
\ead{xiaominji@nju.edu.cn}

\cortext[cor1]{Corresponding authors: Ke-ke Shang (email: kekeshang@nju.edu.cn) and Xiaomin Ji (email: xiaominji@nju.edu.cn).}

\address[a]{Computational Communication Collaboratory, Nanjing University, Nanjing, 210023, China}
\address[b]{School of Journalism and Communication, Nanjing University, Nanjing 210023, China}

\begin{frontmatter}
\begin{abstract}
As artificial intelligence-generated content (AIGC) reshapes knowledge acquisition, higher education faces growing inequities that demand systematic mapping and intervention. We map the AI divide in undergraduate education by combining network science with survey evidence from 301 students at Nanjing University, one of China's leading institutions in AI education. Drawing on course enrolment patterns to construct a disciplinary network, we identify four distinct student communities: science dominant, science peripheral, social sciences \& science, and humanities and social sciences. Survey results reveal significant disparities in AIGC literacy and motivational efficacy, with science dominant students outperforming humanities and social sciences peers. Ordinary least squares (OLS) regression shows that motivational efficacy—particularly skill efficacy—partially mediates this gap, whereas usage efficacy does not mediate at the evaluation level, indicating a dissociation between perceived utility and critical engagement. Our findings demonstrate that curriculum structure and cross-disciplinary integration are key determinants of technological fluency. This work provides a scalable framework for diagnosing and addressing the AI divide through institutional design.

\end{abstract}

\begin{keyword}

AI divide \sep complex networks \sep learning communities \sep interdisciplinary projects \sep motivational efficacy

\end{keyword}

\end{frontmatter}



\section{Introduction}
\label{introduction}
In an era where artificial intelligence is reshaping knowledge production, education is transitioning from knowledge transmission to intelligent collaboration \citep{HAN2025}. Artificial Intelligence Generated Content (AIGC) is a technology that automatically generates multimodal content and is driving this transformation at an unprecedented pace. From the rise of ChatGPT to breakthroughs in models like Deepseek, AIGC has redefined the boundaries of AI applications in education and opened new pathways for personalised learning and higher-order cognitive development.

Against this backdrop, information literacy \citep{VALERI2025100360} has evolved in the AIGC era to encompass not only traditional information acquisition and management capabilities, but also the ability to evaluate the value and risks of AI-generated content. In the context of the AI divide, disparities manifest in three dimensions: access, usage and evaluation. As a user-driven technology, AIGC emphasises the importance of usage and evaluation capabilities \citep{LYTHREATIS2022121359}.

However, existing research has two notable gaps. Firstly, it has not clarified the link between motivational efficacy and AIGC literacy. Secondly, it has not systematically examined the transmission pathway of the education environment - motivation - literacy. Social constructivist theory suggests that learning communities formed through curriculum associations may shape individual motivation via group norms, thereby influencing AIGC literacy. However, the manner in which communities defined by curriculum similarity structure this relationship remains an unexplored area of theory.

As one of China's leading research universities, Nanjing University is a member of the C9 League and plays a pivotal role in the national "Double First-Class" initiative. The university has long been renowned for its commitment to educational innovation and academic excellence.

Nanjing University won the Special Award in the selection of the 7th National Teaching Achievement Award for Higher Education with its project "Exploring the Path of Cultivating Top Innovative Talents in Comprehensive Universities" in September 2014. In September 2023, relying on its profound educational accumulation, Nanjing University launched the undergraduate AI curriculum reform and became the first university in China to build a full-chain curriculum system - from basic AI general courses for undergraduates, to AI interdisciplinary integration courses, and then to core courses focusing on cutting-edge AI technologies.

As a representative university in undergraduate education in China, Nanjing University has a complete range of disciplines and aims to run the best undergraduate education in China. This initiative is a valuable case study for researching curriculum restructuring and content development in higher education. Its systematic approach has also created an ideal environment for analysing the relationship between curriculum and AIGC literacy.

To address these gaps, this study employs community detection and questionnaire surveys to connect meso-level curriculum design with micro-level student literacy at Nanjing University. The research centers on two core questions: How are motivational efficacy and AIGC literacy manifested within communities identified through curriculum similarity? And how do such communities influence AIGC literacy, with what mediating role does motivational efficacy play?

In theory, this research focuses on mapping the environment-motivation-literacy mechanism, offering a new perspective on the formation of the AI divide. In practice, it enables universities to establish dynamic monitoring systems to track students' AIGC literacy development across different curricula, providing data to inform teaching adjustments. By doing so, it promotes the digital transformation of education, advancing it from empirical exploration to data-driven, precise practice.

\section{Literature Review} 

\subsection{AIGC and information literacy}
The rise of AIGC challenges the very foundations of learning\citep{aimag.v26i4.1848}. Education is undergoing a pivotal shift from knowledge transmission to intelligent collaboration – a transformation significantly driven by the rise of AIGC (Artificial Intelligence Generated Content). Artificial intelligence generative adversarial (AIGC) is defined as a production method that leverages AI technology to automatically generate digital content across a variety of media types, including text, images, audio, and video . It has evolved at an unprecedented pace, from the widespread attention initially sparked by ChatGPT 3.5 to the continuous breakthroughs of models like ChatGPT 4.0 and Deepseek \citep{ALQAHTANI20231236}. In contradistinction to conventional AI applications in the field of education, its adaptable and intelligent interactive capabilities in teaching and learning scenarios offer novel pathways for the advancement of personalised education and the cultivation of higher-order cognitive skills \citep{cerny2024ailiteracy}\citep{bazargani2025survey}\citep{dibek2024influence}.

Information literacy is the ability to use digital media to access, create, manage and critically evaluate information and to use it effectively for one's own purposes \citep{LAU201579}. Concurrently, the concept of information literacy has undergone a significant expansion in its scope and complexity\citep{brata2022student}. This evolution is particularly pronounced in the context of the AIGC era, which has introduced novel challenges and expectations, underscoring the need for a multifaceted approach to the development of information literacy \citep{VALERI2025100360}. Education now faces the imperative of equipping students not only with practical skills to operate AIGC tools (e.g., using AI for text creation or data analysis) but also with the capacity to evaluate the value and identify risks in AI-generated content (e.g., assessing credibility and ethical compliance). This will empower them to become "active masters" rather than passive users of intelligent technology.

This urgency is further emphasised by scholars \citep{2bmr-g224-22}, whose review of AI integration in K-12 education highlights the growing adoption of AIGC across grades and disciplines. From an educational equity perspective, AIGC literacy emerges as a critical lever to bridge generational gaps in technology adoption and promote balanced access to educational resources. The fundamental objective of this initiative is to empower individuals to master and ethically utilise intelligent tools and content value, thereby ensuring that technology serves as an engine for inclusive education rather than a catalyst for disparity. This, in turn, establishes the fundamental logic for education to respond to AIGC-driven transformations and cultivate innovative talent.

\subsection{AI divide}
The connotation of the digital divide has evolved from an early binary classification of 'having or lacking access to computers and the internet' to a superimposed pattern of three layers: the access divide (1.0) \citep{dewan2005digital}, the literacy divide (2.0)  \cite{scheerder2017determinants} and the AI divide (3.0) \citep{soomro2020digital} . In the 3.0 stage of the digital divide, the differences in the digital divide are reflected in three dimensions: access, use and evaluation \citep{scheerder2017determinants}.
Artificial Intelligence Generated Content (AIGC) is a user-initiated intelligent technology that provides individuals or groups with the opportunity to actively express themselves and create content. It focuses on differences among individuals or groups in terms of access to and use of intelligent technologies, as well as the results of their application. Its core value lies in the dimensions of use and evaluation. \citep{LYTHREATIS2022121359}
From the use perspective, AIGC literacy emphasises the need for users to be able to express themselves and create content actively with the help of AIGC tools \citep{JIANG2024100287}. From the evaluation dimension, the complexity and potential risks of AIGC-generated content mean that users must be able to evaluate content quality and the boundaries of technical application to effectively avoid risks and apply technology reasonably.

\subsection{Motivation efficacy}
Evolving from traditional ICT literacy, AIGC literacy demands more than just basic operational skills from users; it also requires them to be able to discern, regulate and creatively apply generated content. motivational efficacy, a key variable explaining individual differences in technology application \citep{9694457}, plays an irreplaceable role in shaping AIGC literacy, as evidenced by the research trajectory of ICT literacy. Its influence is amplified by the unique characteristics of AIGC technology, making it even more pronounced.

Motivation efficacy specifically refers to the internal drive that propels individuals to actively explore, learn and continuously apply technological tools \citep{Fraillon2014}. In the ICT domain, this factor has been proven to significantly affect how well and deeply individuals master and apply technology. For example, many young people use ICT tools for socialising and entertainment, but lack the motivation to use them for learning and work, which leaves their related literacy at a superficial level \citep{ROHATGI2016103}. The intensity of motivational efficacy is a key reason for such disparities.

In the realm of AIGC, the importance of motivational efficacy becomes even more apparent. Given the rapid iteration and wide-ranging application scenarios of AIGC technology, users must sustain enthusiasm for continuous learning and recognise its value \citep{GUO2024102547}. Only through frequent technical training can they gain an in-depth understanding of the technology’s logic and effectively apply it to learning contexts. Currently, the connection between motivational efficacy and AIGC literacy has not been specifically explored in academia: this gap serves as the starting point of this paper.

\subsection{Communities detection}
The impact of the environment on individual motivation is well documented. Ecological systems theory emphasises how microenvironments influence the intensity and direction of motivation through resource availability and interaction patterns, while social cognitive theory highlights that supportive environments strengthen goal-oriented motivation by enhancing self-efficacy. In higher education, curriculum systems shape students' cognitive structures \citep{FLIERL201830} through knowledge transmission and construct distinct learning ecosystems through credit requirements, practical opportunities, and disciplinary culture \citep{loh2025plugging}. All of these factors influence students' motivation to adopt AIGC.

Scholars have investigated the relationship between educational environments and the motivational efficacy of AIGC \citep{SCHWEDER2024101829}. Disciplinary curricula differ markedly in their application of AIGC: STEM courses often position AIGC as a data processing tool, with embedded programming and algorithmic exercises that improve operational confidence and promote tool efficacy-driven motivation \citep{VALERI2025100360}. By contrast, humanities curricula focus on the use of AIGC in text generation and creative design \citep{LIU2024104977}. This leads students to prioritise output quality and develop result quality-driven motivation. According to social constructivist theory, interactions within these communities strengthen shared cognition and behaviors, thereby influencing individual exploratory motivation via group norms \citep{SENKBEIL2017145}. In particular, the impact of curricula is amplified through communities, which refers to groups formed by curricular alignment rather than traditional interest-based organizations.

\begin{figure}[ht]
    \centering
    \includegraphics[width=1.0\textwidth]{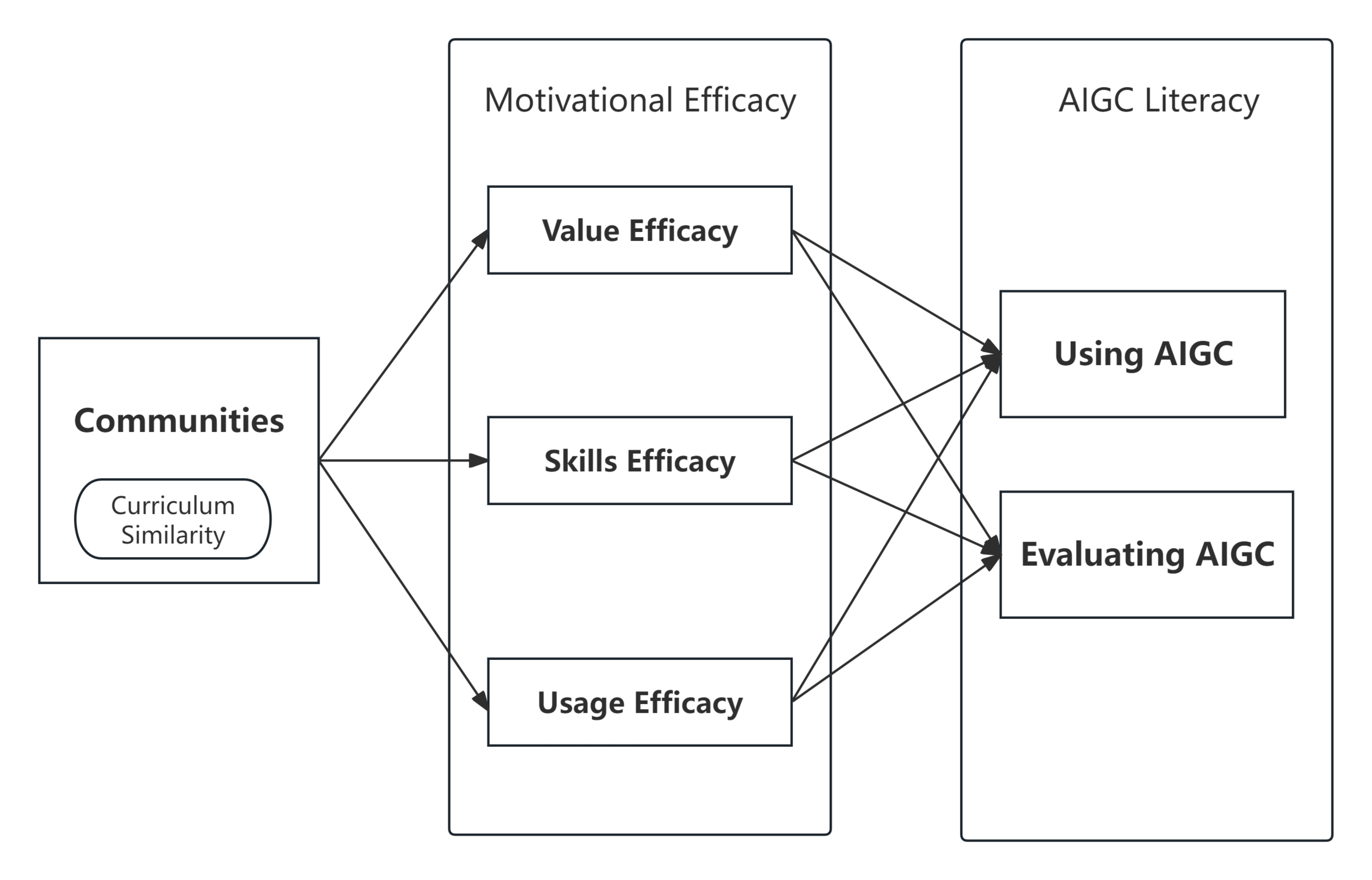}
    \caption{Hypothetical model}
    \label{fig:1}
\end{figure}

Existing studies lack an analysis of the path from educational environmental factors to internal motivation and, subsequently, to AIGC literacy. This is particularly true with regard to the differences in motivational efficacy and literacy development that may be caused by variations in course content. Therefore, this paper proposes the following two research questions (see in figure \ref{fig:1}): 

RQ1: What are the manifestations of motivational efficacy and AIGC literacy among communities formed based on curriculum similarity? 

RQ2: How do communities formed based on curriculum similarity affect students' AIGC literacy, and how does motivational efficacy mediate this effect?

\section{Method}
\subsection{Community detection and measurement}
\subsubsection {Community detection}
In complex networks, a community is defined as a set of nodes that have dense internal connections, but sparse connections to nodes outside the group \citep{GIRVAN20027821}. To identify such structures in our constructed school network, we use the greedy modularity community algorithm. The core logic of this algorithm is to optimise modularity, a metric that measures the quality of community detection \citep{NEWMAN2004026113} \citep{SHANG202068002}. Higher values of modularity indicate stronger within-community connections and weaker inter-community ties. The algorithm begins by treating each node as a separate community and then iteratively merges them to maximise modularity gains until no further improvement is possible. This process classifies schools into communities based on curriculum association characteristics, grouping those with closely related course systems to visualise differences in curriculum structure, thus addressing RQ1.

H1: Students from schools within the same community show similar performance in (a) motivational efficacy and (b) AIGC literacy.

H2: Students from schools in different communities demonstrate different levels of performance in terms of (a) motivational efficacy and (b) AIGC literacy.

\subsubsection {Community measurement}
\begin{figure}[!ht]
    \centering
    \includegraphics[width=1.0\textwidth]{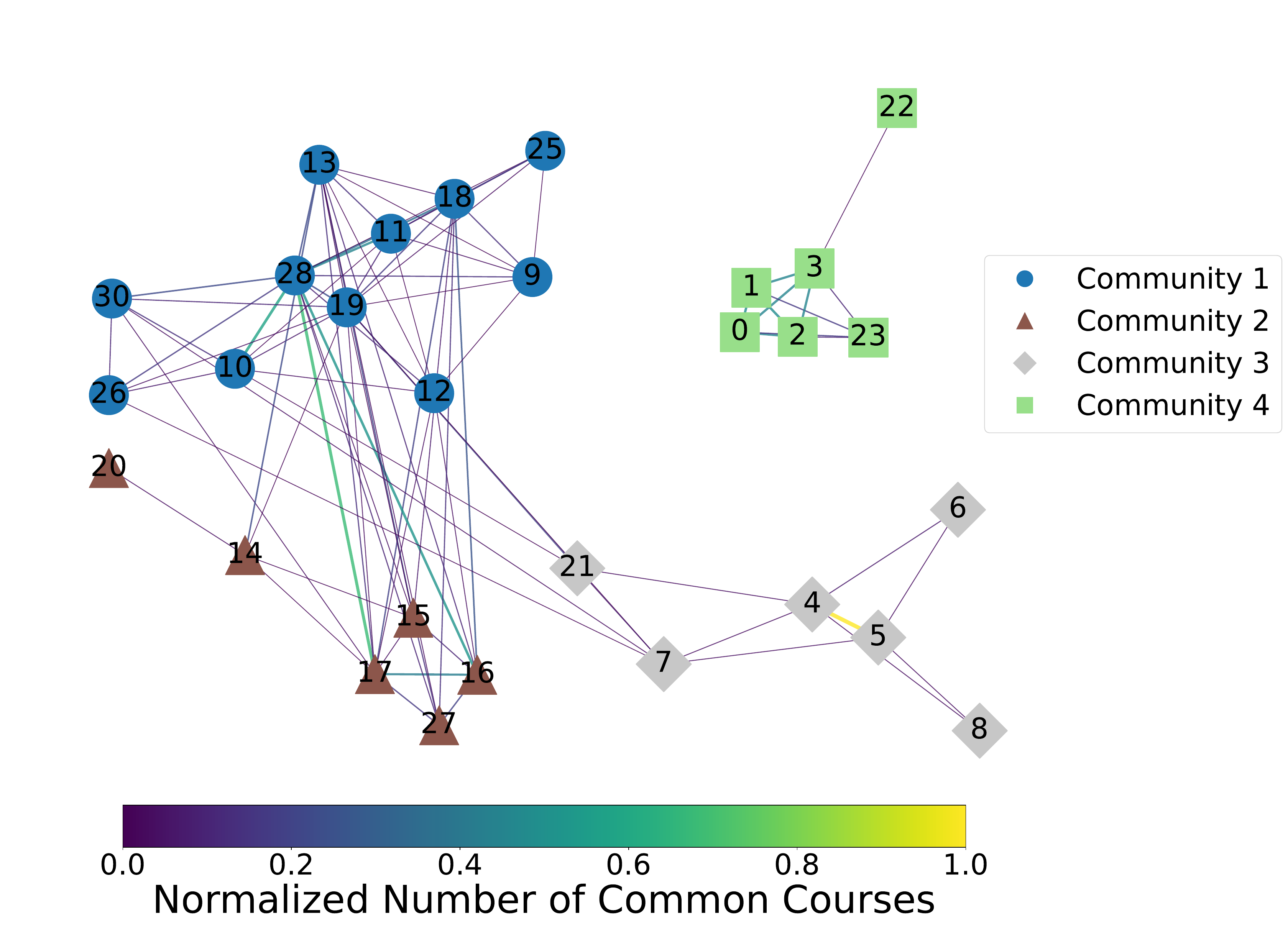}
    \caption{Nodes represent various schools of Nanjing University, and the edges denote curriculum similarity. Four communities are divided based on the characteristics of each school, with the representative school of each community being the one with the highest degree centrality within that community. Among them, Community 1 is a science-dominant community, represented by the Kuang Yaming Honors School; Community 2 is a science-peripheral community, represented by the School of Life Sciences; Community 3 is a science \& social sciences community, represented by the School of Information Management; Community 4 is a humanities and social sciences community, represented by the School of International Studies.. The degree centrality of the four communities decreases sequentially from 1 to 4.}
    \label{fig:2}
\end{figure}

This study uses 31 departments at Nanjing University as nodes in the network, considering the common courses between schools as edges, to explore the association characteristics between shcools. After a comprehensive comparison of indicators such as modularity, coverage and performance score, the community division scheme with a threshold of 3 was selected (modularity 0.427, coverage 0.727 and performance score 0.935), including a total of 29 departments in the disciplinary networks (see in Figure \ref{fig:2}).

\subsection{Variable measurement and questionnaire collection}
\subsubsection {Motivational efficacy measurement}
The present study draws on the research by Martin Senkbeil and Jan Marten Ihme, operationalizes motivational efficacy as a comprehensive manifestation of motivational orientations and efficacy perceptions exhibited by individuals in the use of AIGC. This is measured through three dimensions. These include value efficacy, skill efficacy, and usage efficacy. The value efficacy dimension, corresponding to "motivational access" in the original literature, focuses on individuals' recognition of AIGC's value and reflects their subjective judgment on the value of AIGC. The skill efficacy dimension, derived from the conceptual extension of "Informational Skills Access" in the literature, embodies their perceived ability in information processing and tool operation. The usage efficacy, echoing the core connotation of "Educ Inf Technol" in the original literature, primarily concerns individuals' experience of operational fluency and convenience when actually using AIGC. The measurement of these three dimensions is quantitative in nature, with 3, 5, and 4 items allocated for the measurement of each dimension, respectively.

\subsubsection {AIGC literacy measurement}
In terms of AIGC literacy, this study draws on the measurement approaches of AI literacy due to the scarcity of measurement tools specifically designed for AIGC literacy. Specific reference is made to the Meta AI Literacy Scale (MAILS) \citep{CAROLUS2023100014}. AIGC literacy is operationalized through two core dimensions: using and evaluating AIGC. Using AIGC dimension corresponds to 'AI problem-solving' \citep{Ajzen1985} and focuses on individuals' ability to use AIGC tools to solve practical problems. The 'evaluating AIGC' dimension aligns with the 'AI learning' dimension \citep{CAROLUS2023100114} and emphasises individuals' ability to update their knowledge of AIGC technologies and applications.

Similarly, RQ2 can be addressed:

H3: Motivational efficacy plays a mediating role between community detection and the two core dimensions of AIGC literacy: using and evaluating AIGC.

(a) Motivational efficacy mediates between community detection and the use of AIGC.

(b) Motivational efficacy mediates between community detection and evaluating AIGC.

\subsubsection {Questionnaire data collection}
The questionnaire survey was conducted from March to May 2025. The questionnaires were distributed and collected via Wenjuanxing, a popular questionnaire collection platform in China. The research subjects were undergraduates from the 2021–2024 cohorts at Nanjing University. A total of 310 questionnaires were returned, covering 30 of the university's schools. After excluding nine samples (those from the School of Foreign Languages and the School of Software, which were identified as isolated nodes in the community detection, and those that failed the questionnaire quality control questions), 301 valid questionnaires were collected, giving an effective response rate of $97.1\%$. The respondents consist of 185 males (61.46\%) and 116 females (38.54\%); and cover Nanjing University undergraduate students from the 2021 cohort (49 students, 16.28\%), 2022 cohort (111 students, 36.88\%), 2023 cohort (116 students, 38.54\%), and 2024 cohort (25 students, 8.31\%), and school information is shown in Table \ref{tab:demographics}. For reliability, the Cronbach’s $\alpha$ coefficients of each dimension ranged from $0.766$ to $0.846$ in \ref{tab:r}, indicating good reliability of the questionnaire. As for validity, confirmatory factor analysis showed that $\chi^2$/df $= 1.788 < 3$, GFI $= 0.926 > 0.9$, RMSEA $= 0.052 < 0.1$ in Table \ref{tab:fit}, and all other fitting indices met the standards, demonstrating that the questionnaire had reasonable construct validity.

\begin{table}[htbp]
    \centering
    \caption{School Information of Respondents}
    \begin{tabularx}{\linewidth}{>{\raggedright\arraybackslash}X c c} 
        \toprule
        Variable & Frequency & Percentage (\%) \\
        \midrule
        \hspace{1em}School of Philosophy & 9 & 2.99 \\
        \hspace{1em}School of History & 12 & 3.99 \\
        \hspace{1em}School of Journalism \& Communication & 23 & 7.64 \\
        \hspace{1em}Institute for International Students & 9 & 2.99 \\
        \hspace{1em}School of Government & 11 & 3.65 \\
        \hspace{1em}School of International Studies & 10 & 3.32 \\
        \hspace{1em}Law School & 8 & 2.66 \\
        \hspace{1em}School of Information Management & 9 & 2.99 \\
        \hspace{1em}School of Social and Behavioral Sciences & 5 & 1.66 \\
        \hspace{1em}School of Physics & 6 & 1.99 \\
        \hspace{1em}School of Mathematics & 3 & 1.00 \\
        \hspace{1em}School of Astronomy \& Space Science & 2 & 0.66 \\
        \hspace{1em}School of Atmospheric Sciences & 5 & 1.66 \\
        \hspace{1em}School of Earth Sciences and Engineering & 28 & 9.30 \\
        \hspace{1em}School of Geography and Ocean Sciences & 9 & 2.99 \\
        \hspace{1em}School of Environment & 10 & 3.32 \\
        \hspace{1em}School of Chemistry and Chemical Engineering & 6 & 1.99 \\
        \hspace{1em}School of Life Sciences & 9 & 2.99 \\
        \hspace{1em}school of Engineering and Applied Sciences & 41 & 13.62 \\
        \hspace{1em}School of Management \& Engineering & 2 & 0.66 \\
        \hspace{1em}School of Architecture and Urban Planning & 5 & 1.66 \\
        \hspace{1em}Business School & 7 & 2.33 \\
        \hspace{1em}School of Liberal Arts & 5 & 1.66 \\
        \hspace{1em}School of Marxism & 2 & 0.66 \\
        \hspace{1em}School of Electronic Science and Engineering & 8 & 2.66 \\
        \hspace{1em}School of Computer Science & 16 & 5.32 \\
        \hspace{1em}Medicine School & 16 & 5.32 \\
        \hspace{1em}Kuang Yaming Honors School & 7 & 2.33 \\
        \hspace{1em}School of Artificial Intelligence & 18 & 5.98 \\
        \addlinespace
        \textbf{Total} & 301 & 100.00 \\
        \bottomrule
    \end{tabularx}
    \label{tab:demographics}
\end{table}

\section{Results}
\subsection{Disciplinary networks}
Degree centrality is a measure of the number of direct connections a node has with other nodes. The higher a school's degree centrality, the more courses it shares with other schools, and the more frequently its courses interact. When ranked by degree centrality (see in \ref{tab:degree}) from highest to lowest and combined with the nature of the schools within each community, the four communities in Figure \ref{fig:6} are detected as follows:

\begin{figure}[!htbp]
    \centering
    \includegraphics[width=1.0\textwidth]{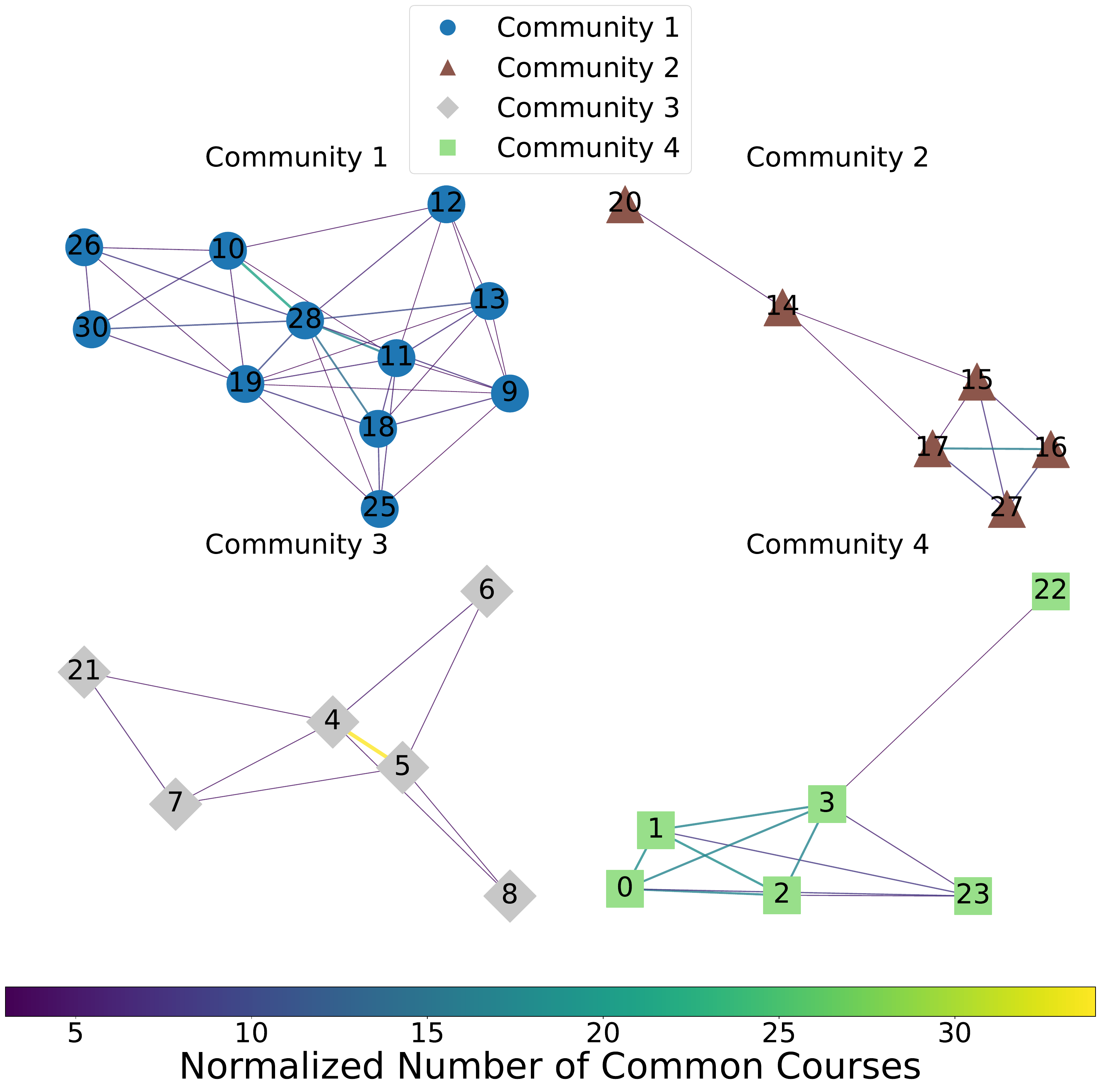}
    \caption{Four communities are divided based on different schools at Nanjing University. Community 1 is a science-dominant community, Community 2 is a science-peripheral community, Community 3 is a science \& social sciences community, and Community 4 is a humanities and social sciences community. The degree centrality of the communities decreases sequentially from 1 to 4.}
    \label{fig:6}
\end{figure}

Community 1 is a science-dominated community. As the community with the highest degree of centrality, it centres on traditional science and engineering departments, such as Kuang Yaming Honors School and School of Engineering Management. It also includes other science and engineering schools, such as the School of Earth Sciences and Engineering and the School of Atmospheric Sciences. These schools excel in subjects with strong logical foundations, such as mathematics, physics and astronomy, boasting intensive course interactions and an extremely high internal correlation. Kuang Yaming Honors School and the School of Engineering Management are the most tightly connected hubs in the disciplinary networks.

Community 2 is a science-peripheral community. With the second-highest degree centrality, it consists of schools like those of life sciences, environment, and chemistry, covering departments such as the School of Chemistry and Chemical Engineering and the School of Medicine. Though falling under the science category, they feature stronger disciplinary professionalism, with lower course correlation than the science-dominated community, thus occupying a peripheral position in the science and engineering networks.

Community 3 is a science \& social sciences community. Ranking third in terms of degree centrality, it takes the School of Information Management and the School of Government Management as its core, bringing together schools like the Business School and the School of International Relations. Its disciplines have both social sciences and science characteristics: information management integrates technology and management, while government management involves social sciences and policies, embodying the nature of interdisciplinary course interactions.

Community 4 is an humanities and social sciences community. With the lowest degree centrality, it comprises humanities and social sciences departments such as the Department of Philosophy, the School of History, and the School of Journalism and Communication. It forms a self-contained system in the field of humanities and social sciences, with close internal course connections. However, it has very few course interactions with other schools, especially those of science and engineering, forming an humanities closed loop.

\subsection{Communities detection and motivation efficacy}
\subsubsection {Internal analysis of communities}
Examining the motivational efficacy of students from different schools within the same community reveals no significant differences in skill and usage efficacy in table \ref{tab:10} \ref{tab:11}. This indicates high consistency in their perceptions of these two efficacies and confirms the rationality of the current community detection in this respect.

However, Value Efficacy is a special case. Significant differences were found between students from different schools in Community 3 (science \& social sciences) in table \ref{tab:12} and Community 4 (humanities and social sciences) in table \ref{tab:13}. Value efficacy is more susceptible to the combined influence of factors such as students' disciplinary backgrounds, their schools' educational philosophies and student group characteristics. This leads to divergence within the same community. Within Community 3, the School of Social and Behavioural Sciences has the highest mean value efficacy ($4.40 \pm 0.28$), which is significantly higher than that of the School of Information Management ($3.52 \pm 0.44$). This may be due to differing perceptions of community value; for instance, social and behavioural science schools may emphasise the community's role in social interaction and identity formation, whereas the School of Information Management may prioritise its practical or instrumental value. In Community 4, the Institute for International Students scores higher, while the School of Journalism \& Communication and the School of Marxism (with a relatively low mean) score lower. This may be due to the different educational goals and focus areas of humanities and social sciences, which are ultimately reflected in their perception of Value Efficacy.

Overall, H1a is partially valid. While the hypothesis that 'students from different schools within the same community show similar motivational efficacy' is supported in terms of skill and usage efficacy, it is not fully valid due to significant differences in value efficacy within communities 3 and 4.

In terms of AIGC literacy in table \ref{tab:14} \ref{tab:15} \ref{tab:16} \ref{tab:17}, the community classification results demonstrate good rationality: students from different schools within the same community show no significant differences in the two core AIGC literacy dimensions of Using AI and Evaluating AI. These results reaffirm the scientific validity of the community classification criteria and indicate that students from various schools within the four communities share a high degree of homogeneity in terms of AIGC literacy. This confirms the validity of Hypothesis H1b.

\subsubsection {External Analysis among communities}
To test hypotheses H2a and H2b, a one-way ANOVA was conducted to examine differences in motivational efficacy and AIGC literacy among the four communities.
The results presented in Table \ref{tab:1} support H2a; significant differences were observed in the three dimensions of motivational efficacy (value efficacy, skill efficacy and usage efficacy) between the different communities. Specifically, the science-dominated community scored highest in all three dimensions (value efficacy: $4.45 \pm 0.50$; skill efficacy: $4.05 \pm 0.53$; usage efficacy: $4.33 \pm0.48$), while the humanities and social sciences community scored lowest. Additionally, the scores of the four communities showed a consistent downward trend, from highest to lowest, in the following order: science-dominated, science-peripheral, social sciences \& science and humanities and social sciences. This further validates the initial community classification criteria, confirming that differences in disciplinary attributes and educational models have shaped students' motivational efficacy characteristics across communities.

\begin{table}[htbp]
    \centering
    \caption{Motivational efficacy analysis among communities}
    \begin{tabularx}{\linewidth}{>{\raggedright\arraybackslash}Xccc}
        \toprule
        Community (M$\pm$SD) & Value Efficacy & Skill Efficacy & Usage Efficacy \\
        \midrule
        science-dominated & 4.45$\pm$0.50 & 4.05$\pm$0.53 & 4.33$\pm$0.48 \\
        science-peripheral & 4.28$\pm$0.44 & 3.88$\pm$0.50 & 4.05$\pm$0.46 \\
        social sciences \& science & 3.95$\pm$0.47 & 3.60$\pm$0.35 & 3.90$\pm$0.34 \\
        humanities and social sciences & 3.81$\pm$0.63 & 3.25$\pm$0.61 & 3.71$\pm$0.47 \\
        F & 26.550 & 36.724 & 29.615 \\
        p & 0.000** & 0.000** & 0.000** \\
        \bottomrule
        \multicolumn{4}{l}{\scriptsize *$p<0.05$ **$p<0.01$} \\
    \end{tabularx}
    \label{tab:1}
\end{table}

The validation results for H2b are consistent with the aforementioned findings (see Table \ref{tab:2}): significant disparities were also observed in AIGC literacy (using and evaluating AIGC) across the different communities. Students in the science-dominated community attained the highest scores in both Using AIGC($4.05 \pm 0.53$) and Evaluating AIGC ($4.33 \pm 0.48$), while those in the humanities and social sciences community scored the lowest (Using AIGC: $3.25 \pm 0.61$; Evaluating AIGC: $3.71 \pm 0.47$). The science-peripheral and social science \& science communities occupied intermediate positions in this ranking. This consistent pattern suggests that the disciplinary curriculum frameworks underpinning each community may be a critical determinant; curricula in the science-dominated community emphasise integration with technical tools, facilitating more frequent and in-depth engagement with AIGC-related knowledge among students. Conversely, the humanities and social sciences community's curricula contain a smaller proportion of content focused on technical applications. These structural discrepancies directly contribute to the observed variations in students' AIGC literacy, thereby validating Hypothesis H2b, and, by extension, Hypothesis H2 as a whole.

\begin{table}[htbp]
    \centering
    \caption{AIGC literacy analysis among communities}
    \begin{tabularx}{\linewidth}{>{\raggedright\arraybackslash}Xcc}
        \toprule
        Community (M$\pm$SD) & Using AI & Evaluating AI \\
        \midrule
        science-dominated & 4.05$\pm$0.53 & 4.33$\pm$0.48 \\
        science-peripheral & 3.88$\pm$0.50 & 4.05$\pm$0.46 \\
        social sciences \& science & 3.60$\pm$0.35 & 3.90$\pm$0.34 \\
        humanities and social sciences & 3.25$\pm$0.61 & 3.71$\pm$0.47 \\
        F & 36.724 & 29.615 \\
        p & 0.000** & 0.000** \\
        \bottomrule
        \multicolumn{3}{l}{\scriptsize *$p<0.05$ **$p<0.01$} \\
    \end{tabularx}
    \label{tab:2}
\end{table}

\subsection{The mediating role of motivational efficacy}

This study employed the Bootstrap test (with 5000 repetitions) to examine the mediating role of motivational efficacy, and determined the significance of the effect by calculating the confidence interval.

\subsubsection{The mediating role between communities and using AIGC}
The data in Table \ref{tab:4} show that the total effect of community detection on students' ability to use AIGC is -0.316. This indicates that, as the correlation with courses decreases and communities shift from science to humanities and social sciences, there is a significant negative change in students' ability to use AIGC. After incorporating motivational efficacy, the direct effect decreases to -0.140, suggesting that motivational efficacy partially mediates this effect. Specifically, the indirect effect through value efficacy is -0.035, reflecting the fact that the shift from science communities to humanities and social sciences communities weakens students' motivation to obtain AIGC information, thereby reducing their ability to use it. The indirect effect through skill efficacy is -0.089, the highest value, indicating that this reduction in technical training directly inhibits efficient AIGC usage. The indirect effect through usage efficacy is -0.052, showing that community detection indirectly leads to insufficient usage efficacy by reducing operational confidence. These results verify H3a, i.e. that motivational efficacy mediates the relationship between community detection and AIGC usage.

\begin{table}[htbp]
    \centering
    \caption{Mediating Effects between Community and Using AIGC}
    \label{tab:4}
    \footnotesize 
    \setlength{\tabcolsep}{5pt}  
    \begin{tabularx}{\textwidth}{>{\raggedright\arraybackslash}X c c c}
        \toprule
        Mediation Path & 
        Total effect (95\% CI) & 
        Direct effect (95\% CI) & 
        Indirect effect (95\% CI) \\
        \midrule
        \makecell[l]{Community $\rightarrow$ VE \\ $\rightarrow$ Using AIGC} 
            & $-0.316\ [-0.367, -0.266]$ 
            & $-0.140\ [-0.191, -0.090]$ 
            & $-0.035\ [-0.124, -0.006]$ \\
        \addlinespace[0.5ex]  
        \makecell[l]{Community $\rightarrow$ SE \\ $\rightarrow$ Using AIGC} 
            & $-0.316\ [-0.367, -0.266]$ 
            & $-0.140\ [-0.191, -0.090]$ 
            & $-0.089\ [-0.239, -0.092]$ \\
        \addlinespace[0.5ex]
        \makecell[l]{Community $\rightarrow$ UE \\ $\rightarrow$ Using AIGC} 
            & $-0.316\ [-0.367, -0.266]$ 
            & $-0.140\ [-0.191, -0.090]$ 
            & $-0.052\ [-0.172, -0.032]$ \\
        \bottomrule
    \end{tabularx}

    \vspace{0.5em}
    \scriptsize Note: VE = Value Efficacy; SE = Skill Efficacy; UE = Usage Efficacy
\end{table}

\subsubsection{The mediating role between communities and evaluating AIGC}
For H3b, the analysis in Table \ref{tab:5} shows that the total effect of community detection on AIGC evaluation ability is -0.255. This indicates that shifting communities towards humanities and social sciences would have a significant negative impact on evaluation ability. After incorporating the three dimensions of motivational efficacy, the direct effect decreased to -0.091, confirming its partial mediating role. The indirect effect through value efficacy is -0.045, indicating that the shift from science to humanities and social sciences reduces students' initiative to obtain AIGC information, thereby indirectly weakening their ability to evaluate the limitations and opportunities of AI. The indirect effect through skill efficacy is -0.096, the highest value, reflecting that a reduction in technical training directly inhibits the ability to make critical judgements about the advantages and disadvantages of AI. Notably, the mediating effect of usage efficacy is not established, showing that operational confidence is not significantly correlated with high-level thinking, such as evaluating the ethical risks and innovative possibilities of AI. Thus, evaluation ability depends more on information analysis and logical judgement skills than on simple technical operations. Together, value efficacy and skill efficacy explain 55.46 \% of the total effect, further verifying H3b.
In conclusion, motivational efficacy plays a mediating role between community detection and the core dimensions of AIGC literacy, and hypothesis H3 is generally valid.

\begin{table}[htbp]
    \centering
    \caption{Mediating Effects between Community and Evaluating AIGC}
    \label{tab:5}
    \footnotesize  
    \setlength{\tabcolsep}{5pt}  
    \begin{tabularx}{\textwidth}{>{\raggedright\arraybackslash}X c c c}
        \toprule
        Mediation Path & 
        Total effect (95\% CI) & 
        Direct effect (95\% CI) & 
        Indirect effect (95\% CI) \\
        \midrule
        \makecell[l]{Community $\rightarrow$ VE \\ $\rightarrow$ Evaluating AIGC} 
            & $-0.255\ [-0.304, -0.206]$ 
            & $-0.091\ [-0.142, -0.041]$ 
            & $-0.045\ [0.094, 0.317]$ \\
        \addlinespace[0.5ex]
        \makecell[l]{Community $\rightarrow$ SE \\ $\rightarrow$ Evaluating AIGC} 
            & $-0.255\ [-0.304, -0.206]$ 
            & $-0.091\ [-0.142, -0.041]$ 
            & $-0.096\ [0.254, 0.483]$ \\
        \addlinespace[0.5ex]
        \makecell[l]{Community $\rightarrow$ UE \\ $\rightarrow$ Evaluating AIGC} 
            & $-0.255\ [-0.304, -0.206]$ 
            & $-0.091\ [-0.142, -0.041]$ 
            & $-0.022\ [-0.036, 0.251]$ \\
        \bottomrule
    \end{tabularx}

    \vspace{0.5em}
    \scriptsize Note: VE = Value Efficacy; SE = Skill Efficacy; UE = Usage Efficacy
\end{table}

\section{Discussion}
The emergence of AIGC technology is reshaping higher education, offering transformative potential for talent development while challenging pedagogical models \citep{aimag.v26i4.1848}. As a leading institution in AI research, Nanjing University has introduced dedicated AIGC-related courses early on. However, systematic development of AIGC literacy remains a critical challenge. Our study introduces an integrative framework combining community network detection with large-scale survey data, aiming to inform evidence-based strategies for educational innovation in the age of generative AI.

Our analysis reveals that the four communities identified through curriculum-based clustering exhibit strong associations between structural organization and student motivational efficacy. Greater curriculum similarity facilitates interdisciplinary content integration, fostering active engagement with AIGC applications \citep{SENKBEIL2017145}. However, value efficacy is less responsive to curricular integration alone, suggesting the need for targeted interdisciplinary modules that preserve foundational disciplinary values while promoting cross-domain collaboration through shared project-based learning.

A comparative assessment of AIGC literacy across disciplinary clusters highlights pronounced disparities. Learners in science-dominant clusters demonstrate superior competence in tool utilization and outcome evaluation, whereas those in humanities and social sciences clusters exhibit lower performance. This divergence underscores the necessity of establishing structured collaboration mechanisms between science and social science education. Two strategies are proposed: interdisciplinary laboratories where STEM learners lead technical development while humanities and social sciences learners contribute to scenario design and ethical assessment, and 'AIGC + discipline' micro-programmes embedding technical application units into humanities curricula and introducing modules on technological ethics into STEM pathways \citep{JIANG2024100287}.

As curriculum relevance declines, community composition shifts from science-dominant to humanities-focused, and the mediating role of motivational efficacy becomes increasingly salient. Structured curricular design plays a pivotal role in strengthening perceived value, skill confidence, and behavioral intention. At the evaluation level, usage efficacy does not mediate, indicating a need for curricular inclusion of modules on ethical analysis and validation of AIGC outputs \citep{loh2025plugging}.

Previous research on the digital divide and ICT literacy has long emphasised access to information as a central determinant \citep{ROHATGI2016103}. While the present study focuses on variation in AIGC usage and evaluation across institutional contexts, it initially omitted the antecedent dimension of information access. To address this gap, access was operationalised as depth and breadth of curricular exposure. An independent samples t-test, stratified by academic cohort (Table \ref{tab:6}), shows that the 2021 cohort achieved the highest scores in both 'Using AIGC' and 'Evaluating AIGC', whereas the 2024 cohort scored lowest, with statistically significant differences (F=5.533, p=0.001**; F=2.940, p=0.033*). This gradient arises from cumulative curricular immersion: the 2021 cohort completed a full undergraduate programme, during which motivational efficacy was systematically reinforced through progressive learning, establishing a robust foundation for literacy. In contrast, the 2024 cohort, having recently entered university, has completed only foundational general education courses, with minimal exposure to advanced disciplinary content. Consequently, the motivational scaffolding provided by curricula remains underdeveloped, constraining the emergence of AIGC literacy.

\begin{table}[htbp]
    \centering
    \caption{Different analysis of admitted class}
    \label{tab:6}
    \begin{tabularx}{\linewidth}{>{\raggedright\arraybackslash}Xcc}
        \toprule
        Admitted class ($M \pm SD$) & Using AI & Evaluating AI \\
        \midrule
        2021 ($n=49$)  & $4.17 \pm 0.62$ & $4.07 \pm 0.60$ \\
        2022 ($n=111$) & $4.05 \pm 0.58$ & $4.02 \pm 0.55$ \\
        2023 ($n=116$) & $3.82 \pm 0.66$ & $3.92 \pm 0.58$ \\
        2024 ($n=25$)  & $3.73 \pm 0.74$ & $3.69 \pm 0.76$ \\
        \addlinespace
        $F$            & 5.533           & 2.940           \\
        $p$            & $0.001^{**}$    & $0.033^{*}$     \\
        \bottomrule
        \multicolumn{3}{l}{\scriptsize *$p<0.05$ **$p<0.01$} \\
    \end{tabularx}
\end{table}

This finding underscores the importance of curricular coherence and longitudinal progression in cultivating AIGC literacy. Rather than limiting integration to advanced courses, institutions should embed AIGC across the learning cycle: introducing foundational concepts in general education to spark engagement, and deepening discipline-specific applications in upper-level courses—such as AI-assisted text generation in the humanities or data analysis in STEM. Our contribution lies in mapping this developmental model, aligning AIGC literacy with cognitive and professional growth, and offering a practical pathway for interdisciplinary talent development in the age of AI.

The interpretation of these findings is constrained by the single-institution design and non-random sampling, which may limit generalizability and introduce selection bias. As the study was conducted at a leading AI-education institution, cultural and pedagogical specificities may influence outcomes. Moreover, while key dimensions of AI literacy are addressed, the construct of information access—how students locate and evaluate AI-related knowledge—remains underdeveloped. Future research should validate these insights across diverse institutions and explicitly examine access as a determinant of equitable AI learning.

\clearpage
\appendix
\section{Reliability Test Results}
\begin{table}[!ht]
\centering
\caption{Reliability Test Results}
\begin{tabularx}{\linewidth}{lXcc}
\toprule
\textbf{Variable} & \textbf{Items} & \textbf{SFL} & \textbf{Cronbach$\alpha$} \\
\midrule
\multirow{4}{*}{\shortstack[l]{Motivation \\ Access}} 
& Using the Internet can provide me with information that would lead to better decisions. & 0.876 & \multirow{4}{*}{0.817} \\
& Using computer and Internet can improve my work performance. & 0.807 &  \\
& Using Computer and the Internet seem to be enjoyable & 0.669 &  \\
& I always know what search terms to use when searching the internet. & 0.644 &  \\
\midrule
\multirow{5}{*}{\shortstack[l]{Skill \\ Access}} 
& I can use advance search options to reach my required information. & 0.788 & \multirow{5}{*}{0.846} \\
& I feel confident to evaluate the sources of the information found on the Internet. & 0.732 &  \\
& I feel well to synthesize online information. & 0.725 &  \\
& It is easy for me to retrieve a Website on the Internet. & 0.736 &  \\
& On the Internet, it is easy for me to work toward a specific goal. & 0.832 &  \\
\midrule
\multirow{5}{*}{\shortstack[l]{Technical \\ Access}} 
& I can gain benefits from using Internet. & 0.631 & \multirow{5}{*}{0.845} \\
& Using various ICT tools, I feel confident in achieving my goals. & 0.834 &  \\
& I feel confident in making important decisions with the help of the Internet. & 0.723 &  \\
& When using AI, I can rely on my skills in difficult situations. & 0.724 &  \\
\midrule
\multirow{3}{*}{\shortstack[l]{Using \\ AIGC}} 
& I can handle most of the problems related to AIGC by myself. & 0.831 & \multirow{3}{*}{0.819} \\
& When using AIGC well, I can usually solve heavy and complex tasks. & 0.786 &  \\
\midrule
\multirow{3}{*}{\shortstack[l]{Evaluating \\ AIGC}} 
& I can assess what the limitations and opportunities of using AIGC. & 0.768 & \multirow{3}{*}{0.766} \\
& I can assess what advantages and disadvantages the use of AIGC entails. & 0.761 &  \\
& I can think of new uses for AIGC & 0.673 &  \\
\bottomrule
\multicolumn{4}{l}{\scriptsize  Note: SFL = Standardized Factor Loading} \\
\end{tabularx}
\label{tab:r} 
\end{table}

\section{Validity Test Results}
\begin{table}[htbp]
    \centering
    \caption{Validity Test Results}
    \begin{tabular}{cccc}
        \hline
        Indicator & Value & Standard & Whether Qualified \\
        \hline
        $\chi^2$ & 223.542 & - & - \\
        $df$ & 125 & - & - \\
        $\chi^2/df$ & 1.788 & $<$ 3 & Yes \\
        $GFI$ & 0.926 & $>$ 0.9 & Yes \\
        $RMSEA$ & 0.052 & $<$ 0.10 & Yes \\
        $RMR$ & 0.022 & $<$ 0.05 & Yes \\
        $CFI$ & 0.965 & $>$ 0.9 & Yes \\
        $NFI$ & 0.925 & $>$ 0.9 & Yes \\
        $NNFI$ & 0.957 & $>$ 0.9 & Yes \\
        $TLI$ & 0.957 & $>$ 0.9 & Yes \\
        $IFI$ & 0.965 & $>$ 0.9 & Yes \\
        $PGFI$ & 0.677 & $>$ 0.5 & Yes \\
        $PNFI$ & 0.755 & $>$ 0.5 & Yes \\
        $PCFI$ & 0.788 & $>$ 0.5 & Yes \\
        $SRMR$ & 0.044 & $<$ 0.1 & Yes \\
        $RMSEA\ 90\%\ CI$ & 0.040 -- 0.062 & - & - \\
        \hline
    \end{tabular}
    \label{tab:fit} 
\end{table}

\section{School Nodes and Degree Centrality}
\begin{table}[!ht]
    \centering
    \caption{Schools are ranked in descending order of their degree centrality values to illustrate the differences in centrality among various schools within disciplinary networks. Here, blue represents Community 1, a science-dominant community; brown represents Community 2, a science-peripheral community; gray represents Community 3, an social sciences \& science community; and green represents Community 4, a humanities and social sciences community.}
    \begin{tabular}{|c|c|c|c|}
    \hline
    \textbf{Node} & \textbf{School} & \textbf{DC} & \textbf{Ranking} \\ \hline
    28 & \cellcolor{customblue}Kuang Yaming Honors School & 0.5 & 1 \\ \hline
    19 & \cellcolor{customblue}School of Management \& Engineering & 0.5 & 2 \\ \hline
    13 & \cellcolor{customblue}School of Earth Sciences and Engineering & 0.392857 & 3 \\ \hline
    17 & \cellcolor{custombrown}School of Life Sciences & 0.357143 & 4 \\ \hline
    18 & \cellcolor{customblue}College of Engineering and Applied Sciences & 0.357143 & 5 \\ \hline
    11 & \cellcolor{customblue}School of Astronomy \& Space Science & 0.285714 & 6 \\ \hline
    15 & \cellcolor{custombrown}School of Environment & 0.285714 & 7 \\ \hline
    10 & \cellcolor{customblue}School of Mathematics & 0.25 & 8 \\ \hline
    16 & \cellcolor{custombrown}School of Chemistry and Chemical Engineering & 0.25 & 9 \\ \hline
    9 & \cellcolor{customblue}School of Physics & 0.25 & 10 \\ \hline
    12 & \cellcolor{customblue}School of Atmospheric Sciences & 0.25 & 11 \\ \hline
    30 & \cellcolor{customblue}School of Artificial Intelligence & 0.214826 & 12 \\ \hline
    27 & \cellcolor{custombrown}Medicine School & 0.214826 & 13 \\ \hline
    7 & \cellcolor{customgray}School of Information Management & 0.214826 & 14 \\ \hline
    4 & \cellcolor{customblue}School of Government & 0.178571 & 15 \\ \hline
    3 & \cellcolor{customgreen}Institute for International Students & 0.178571 & 16 \\ \hline
    14 & \cellcolor{custombrown}School of Geography and Ocean Sciences & 0.178571 & 17 \\ \hline
    25 & \cellcolor{customblue}School of Electronic Science and Engineering & 0.178571 & 18 \\ \hline
    26 & \cellcolor{customblue}School of Computer Science & 0.178571 & 19 \\ \hline
    5 & \cellcolor{customgray}School of International Studies & 0.142857 & 20 \\ \hline
    0 & \cellcolor{customgreen}School of Philosophy & 0.142857 & 21 \\ \hline
    1 & \cellcolor{customgreen}School of History & 0.142857 & 22 \\ \hline
    2 & \cellcolor{customgreen}School of Journalism \& Communication & 0.142857 & 23 \\ \hline
    21 & \cellcolor{customgray}Business School & 0.142857 & 24 \\ \hline
    23 & \cellcolor{customgreen}School of Marxism & 0.142857 & 25 \\ \hline
    6 & \cellcolor{customgray}Law School & 0.071429 & 26 \\ \hline
    8 & \cellcolor{custombrown}School of Social and Behavioral Sciences & 0.071429 & 27 \\ \hline
    20 & \cellcolor{custombrown}School of Architecture and Urban Planning & 0.035714 & 28 \\ \hline
    22 & \cellcolor{customblue}School of Liberal Arts & 0.035714 & 29 \\ \hline
    24 & School of Foreign Languages & 0 & Null \\ \hline
    29 & School of Software & 0 & Null \\ \hline
    \end{tabular}
\label{tab:degree}
\end{table}

\section{Communities of motivation efficacy}

\subsection{Community 1 (science-dominated) of motivation efficacy}
\begin{table}[htbp]
    \centering
    \caption{Community 1 (science-dominated) of motivation efficacy}
    \begin{tabularx}{\linewidth}{>{\raggedright\arraybackslash}Xccc} 
        \toprule
        Community 1 (M$\pm$SD) & Value Efficacy & Skill Efficacy & Usage Efficacy \\
        \midrule
        School of Artificial Intelligence  & 4.56$\pm$0.47 & 3.91$\pm$0.61 & 4.24$\pm$0.47 \\
        Kuang Yaming Honors School  & 4.33$\pm$0.47 & 3.83$\pm$0.34 & 4.29$\pm$0.51 \\
        School of Earth Sciences and Engineering  & 4.30$\pm$0.61 & 3.85$\pm$0.43 & 4.21$\pm$0.44 \\
        School of Atmospheric Sciences  & 4.47$\pm$0.38 & 4.00$\pm$0.37 & 4.20$\pm$0.45 \\
        School of Astronomy \& Space Science  & 5.00$\pm$0.00 & 4.40$\pm$0.85 & 5.00$\pm$0.00 \\
        School of Management \& Engineering  & 4.00$\pm$0.00 & 3.80$\pm$0.28 & 4.00$\pm$0.00 \\
        School of Mathematics  & 4.56$\pm$0.51 & 4.47$\pm$0.76 & 4.58$\pm$0.52 \\
        School of Physics  & 4.61$\pm$0.39 & 4.13$\pm$0.52 & 4.42$\pm$0.56 \\
        College of Engineering and Applied Sciences  & 4.43$\pm$0.46 & 4.18$\pm$0.58 & 4.39$\pm$0.51 \\
        School of Electronic Science and Engineering  & 4.54$\pm$0.43 & 4.03$\pm$0.23 & 4.28$\pm$0.39 \\
        School of Computer Science  & 4.52$\pm$0.53 & 4.22$\pm$0.53 & 4.39$\pm$0.46 \\
        F & 0.907 & 1.433 & 0.970 \\
        p & 0.529 & 0.173 & 0.473 \\
        \bottomrule
        \multicolumn{4}{l}{\scriptsize *$p<0.05$ **$p<0.01$} \\
    \label{tab:10}
    \end{tabularx}
\end{table}

\subsection{Community 2 (science-peripheral) of motivation efficacy}
\begin{table}[htbp]
    \centering
    \caption{Community 2 (science-peripheral) of motivation efficacy}
    \begin{tabularx}{\linewidth}{>{\raggedright\arraybackslash}Xccc}
        \toprule
        Community 2 (M$\pm$SD) & Value Efficacy & Skill Efficacy & Usage Efficacy \\
        \midrule
        School of Chemistry and Chemical Engineering & 4.44$\pm$0.50 & 3.87$\pm$0.99 & 4.17$\pm$0.68 \\
        Medicine School & 4.15$\pm$0.34 & 3.82$\pm$0.30 & 4.00$\pm$0.30 \\
        School of Geography and Ocean Sciences & 4.41$\pm$0.60 & 3.91$\pm$0.35 & 3.86$\pm$0.38 \\
        School of Architecture and Urban Planning & 4.27$\pm$0.37 & 3.60$\pm$0.40 & 4.15$\pm$0.49 \\
        School of Environment & 4.37$\pm$0.43 & 4.00$\pm$0.55 & 4.20$\pm$0.63 \\
        School of Life Sciences & 4.22$\pm$0.44 & 4.00$\pm$0.54 & 4.06$\pm$0.39 \\
        F & 0.715 & 0.548 & 0.662 \\
        p & 0.615 & 0.739 & 0.654 \\
        \bottomrule
        \multicolumn{4}{l}{\scriptsize *$p<0.05$ **$p<0.01$} \\
    \label{tab:11}
    \end{tabularx}
\end{table}

\subsection{Community 3 (social sciences \& science) of motivation efficacy}
\begin{table}[htbp]
    \centering
    \caption{Community 3 (social sciences \& science) of motivation efficacy}
    \begin{tabularx}{\linewidth}{>{\raggedright\arraybackslash}Xccc}
        \toprule
        Community 3 (M$\pm$SD) & Value Efficacy & Skill Efficacy & Usage Efficacy \\
        \midrule
        School of Information Management & 3.52$\pm$0.44 & 3.67$\pm$0.40 & 3.78$\pm$0.26 \\
        Business School & 4.10$\pm$0.42 & 3.66$\pm$0.28 & 4.11$\pm$0.28 \\
        School of International Studies & 4.13$\pm$0.39 & 3.64$\pm$0.39 & 3.88$\pm$0.40 \\
        School of Government & 3.91$\pm$0.34 & 3.49$\pm$0.39 & 3.95$\pm$0.27 \\
        Law School & 3.88$\pm$0.56 & 3.60$\pm$0.39 & 3.81$\pm$0.50 \\
        School of Social and Behavioral Sciences & 4.40$\pm$0.28 & 3.60$\pm$0.14 & 3.90$\pm$0.22 \\
        F & 3.718 & 0.320 & 0.925 \\
        p & 0.007** & 0.898 & 0.474 \\
        \bottomrule
        \multicolumn{4}{l}{\scriptsize *$p<0.05$ **$p<0.01$} \\
    \label{tab:12}
    \end{tabularx}
\end{table}

\subsection{Community 4 (humanities and social sciences) of motivation efficacy}
\begin{table}[htbp]
    \centering
    \caption{Community 4 (humanities and social sciences) of motivation efficacy}
    \begin{tabularx}{\linewidth}{>{\raggedright\arraybackslash}Xccc}
        \toprule
        Community 4 (M$\pm$SD) & Value Efficacy & Skill Efficacy & Usage Efficacy \\
        \midrule
        School of History & 3.92$\pm$0.47 & 3.10$\pm$0.61 & 3.58$\pm$0.60 \\
        School of Philosophy & 4.00$\pm$0.00 & 3.07$\pm$0.49 & 3.69$\pm$0.46 \\
        School of Liberal Arts & 4.07$\pm$0.37 & 3.56$\pm$0.38 & 3.75$\pm$0.43 \\
        School of Journalism \& Communication & 3.56$\pm$0.58 & 3.33$\pm$0.58 & 3.77$\pm$0.48 \\
        Institute for International Students & 4.15$\pm$0.44 & 3.36$\pm$0.37 & 3.72$\pm$0.36 \\
        School of Marxism & 2.50$\pm$2.12 & 2.50$\pm$2.12 & 3.50$\pm$0.71 \\
        F & 4.379 & 1.326 & 0.316 \\
        p & 0.002** & 0.267 & 0.902 \\
        \bottomrule
        \multicolumn{4}{l}{\scriptsize *$p<0.05$ **$p<0.01$} \\
    \label{tab:13}
    \end{tabularx}
\end{table}

\section{Communities of AIGC literacy}
\subsection{Community 1 (science-dominated) of AIGC literacy}
\begin{table}[htbp]
    \centering
    \caption{Community 1 (science-dominated) of AIGC literacy}
    \begin{tabularx}{\linewidth}{>{\raggedright\arraybackslash}Xcc}
        \toprule
        Community 1 (M$\pm$SD) & Using AI & Evaluating AI \\
        \midrule
        School of Artificial Intelligence & 4.31$\pm$0.48 & 4.28$\pm$0.55 \\
        Kuang Yaming Honors School & 4.57$\pm$0.53 & 4.19$\pm$0.79 \\
        School of Earth Sciences and Engineering & 4.21$\pm$0.53 & 4.26$\pm$0.51 \\
        School of Atmospheric Sciences & 4.00$\pm$0.24 & 4.20$\pm$0.30 \\
        School of Astronomy \& Space Science & 4.50$\pm$0.71 & 4.50$\pm$0.71 \\
        School of Management \& Engineering & 3.83$\pm$0.24 & 3.50$\pm$0.71 \\
        School of Mathematics & 4.56$\pm$0.51 & 4.67$\pm$0.58 \\
        School of Physics & 4.17$\pm$0.84 & 4.33$\pm$0.63 \\
        College of Engineering and Applied Sciences & 4.28$\pm$0.50 & 4.20$\pm$0.53 \\
        School of Electronic Science and Engineering & 4.33$\pm$0.40 & 4.13$\pm$0.31 \\
        School of Computer Science & 4.44$\pm$0.61 & 4.42$\pm$0.54 \\
        \midrule
        F & 0.838 & 0.871 \\
        p & 0.593 & 0.562 \\
        \bottomrule
        \multicolumn{3}{l}{\scriptsize *$p<0.05$ **$p<0.01$} \\
    \end{tabularx}
    \label{tab:14}
\end{table}

\subsection{Community 2 (science-peripheral) of AIGC literacy}
\begin{table}[htbp]
    \centering
    \caption{Community 2 (science-peripheral) of AIGC literacy}
    \begin{tabularx}{\linewidth}{>{\raggedright\arraybackslash}Xcc}
        \toprule
        Community 2 (M$\pm$SD) & Using AI & Evaluating AI \\
        \midrule
        School of Chemistry and Chemical Engineering & 4.22$\pm$0.66 & 4.22$\pm$0.62 \\
        Medicine School & 3.92$\pm$0.35 & 3.92$\pm$0.23 \\
        School of Geography and Ocean Sciences & 3.96$\pm$0.48 & 3.96$\pm$0.51 \\
        School of Architecture and Urban Planning & 3.93$\pm$0.60 & 3.60$\pm$0.37 \\
        School of Environment & 4.10$\pm$0.69 & 4.07$\pm$0.72 \\
        School of Life Sciences & 4.07$\pm$0.43 & 3.93$\pm$0.55 \\
        \midrule
        F & 0.430 & 0.964 \\
        p & 0.826 & 0.449 \\
        \bottomrule
        \multicolumn{3}{l}{\scriptsize *$p<0.05$ **$p<0.01$} \\
    \end{tabularx}
    \label{tab:15}
\end{table}

\subsection{Community 3 (social sciences \& science) of AIGC literacy}
\begin{table}[htbp]
    \centering
    \caption{Community 3 (social sciences \& science) of AIGC literacy}
    \begin{tabularx}{\linewidth}{>{\raggedright\arraybackslash}Xcc}
        \toprule
        Community 3 (M$\pm$SD) & Using AI & Evaluating AI \\
        \midrule
        School of Information Management & 3.52$\pm$0.44 & 3.74$\pm$0.40 \\
        Business School & 3.57$\pm$0.42 & 3.71$\pm$0.49 \\
        School of International Studies & 3.60$\pm$0.34 & 3.80$\pm$0.32 \\
        School of Government & 3.82$\pm$0.56 & 3.79$\pm$0.40 \\
        Law School & 4.08$\pm$0.58 & 3.75$\pm$0.30 \\
        School of Social and Behavioral Sciences & 3.67$\pm$0.24 & 3.80$\pm$0.18 \\
        \midrule
        F & 1.703 & 0.072 \\
        p & 0.154 & 0.996 \\
        \bottomrule
        \multicolumn{3}{l}{\scriptsize *$p<0.05$ **$p<0.01$} \\
    \end{tabularx}
    \label{tab:16}
\end{table}

\subsection{Community 4 (humanities and social sciences) of AIGC literacy}
\begin{table}[htbp]
    \centering
    \caption{Community 4 (humanities and social sciences) of AIGC literacy}
    \begin{tabularx}{\linewidth}{>{\raggedright\arraybackslash}Xcc}
        \toprule
        Community 4 (M$\pm$SD) & Using AI & Evaluating AI \\
        \midrule
        School of History & 3.17$\pm$0.50 & 3.39$\pm$0.74 \\
        School of Philosophy & 3.41$\pm$0.55 & 3.44$\pm$0.37 \\
        School of Liberal Arts & 3.20$\pm$0.38 & 3.73$\pm$0.28 \\
        School of Journalism \& Communication & 3.41$\pm$0.53 & 3.57$\pm$0.52 \\
        Institute for International Students & 3.52$\pm$0.58 & 3.48$\pm$0.34 \\
        School of Marxism & 2.33$\pm$1.89 & 2.33$\pm$1.89 \\
        \midrule
        F & 1.716 & 1.965 \\
        p & 0.147 & 0.099 \\
        \bottomrule
        \multicolumn{3}{l}{\scriptsize *$p<0.05$ **$p<0.01$} \\
    \end{tabularx}
    \label{tab:17}
\end{table}

\end{document}